\documentstyle[aps,preprint]{revtex}
\def\laq{\ \raise 0.4ex\hbox{$<$}\kern -0.8em\lower 0.62
ex\hbox{$\sim$}\ }
\def\gaq{\ \raise 0.4ex\hbox{$>$}\kern -0.7em\lower 0.62
ex\hbox{$\sim$}\ }

\def\NPB{{\em Nucl. Phys.} B}
\def\PLB{{\em Phys. Lett.}  B}
\def\PRL{{\em Phys. Rev. Lett. }}
\def\PRD{{\em Phys. Rev.} D}

\def\pbd{{\dot {\bar \phi}}}

\begin{document}


\title{Generalized Second Law in Cosmology 
From Causal Boundary Entropy }

\author{Ram Brustein}
\address{Department of Physics,
Ben-Gurion University,
Beer-Sheva 84105, Israel\\
email: ramyb@bgumail.bgu.ac.il}

\maketitle

\begin{abstract}
A  classical and quantum mechanical generalized second law of thermodynamics in
cosmology  implies constraints on the effective equation of state of the
universe in the form of energy conditions, obeyed by many known cosmological
solutions, forbids certain cosmological singularities, and is compatible with
entropy bounds. This second law is based on the conjecture that causal 
boundaries and not only event horizons have geometric
entropies proportional to their area.
In string cosmology the second law provides new information about  non-singular  
solutions.
 \end{abstract}

\vspace{-.2in}
Cosmological singularities have been  investigated, relying on the
celebrated singularity theorems of Hawking and Penrose \cite{singth}, who
concluded that if sources in Einstein's equations obey certain energy
conditions, cosmological singularities are inevitable. Entropy considerations
were brought in only much later, when Bekenstein \cite{beb1} argued that if the
entropy of a visible part of the universe obeys the usual entropy bound from
nearly flat space situations \cite{beb2}, certain cosmological singularities
are thermodynamically unacceptable. Recently, Veneziano \cite{veb} suggested
that since a black hole larger than a cosmological horizon cannot form
\cite{ch},  the entropy of the universe is always bounded. This
suggestion is related, although not always equivalent, to the application of
the holographic principle \cite{holo} in cosmology \cite{fs,EL,KL,otherhol}.

I propose  a concrete classical and quantum mechanical form of a
generalized second law (GSL) of thermodynamics in cosmology, valid
also in situations far from thermal equilibrium, discuss various
entropy sources, such as thermal, geometric and quantum entropy,
apply GSL to study cosmological solutions, and show that it is
compatible with entropy bounds. GSL allows a more detailed
description of how, and if, cosmological singularities are evaded.
The proposed GSL is different from GSL for black holes
\cite{gslbh}, but the idea that in addition to normal entropy
other sources of entropy have to be included has some
similarities.

That systems with event horizons, such as black holes and a deSitter
universe have entropy proportional to the area of their horizon
is by now an accepted fact. The proposed GSL is based on the (reasonable) 
conjecture  that  causal  boundaries and 
not only event horizons have geometric entropies proportional to their area.
However, since the conjecture has not been proved yet, further  
investigation could reveal that it is 
incorrect or applies only in special situations. 
A proof of the conjecture will put our results on a much firmer ground.

The starting point of our classical discussion is the definition
of the total entropy of a domain containing more than one
cosmological horizon \cite{veb}. For a given scale factor $a(t)$,
and a Hubble parameter $H(t)=\dot a/ a$, the number of
cosmological horizons within a given comoving volume $V=a(t)^3$ is
simply the total volume divided by the volume of a single horizon,
$n_H={a(t)^3}/{ |H(t)|^{-3}}$ (we will ignore numerical factors of order unity,
use units in which $c=1, G_{N}=1/16\pi, \hbar=1$ and discuss only
flat, homogeneous, and isotropic cosmologies). If the entropy
within a given horizon is $S^H$, then the total entropy is given
by $ S=n_H S^H$. Classical GSL  requires that the cosmological
evolution, even when far from thermal equilibrium, must obey
$
 dS \ge 0,
$ in addition to Einstein's equations. In particular,
\begin{equation}
 n_H \partial_t S^H+ \partial_t n_H S^H \ge 0.
\label{myeq}
\end{equation}

In general, there could be many sources and types of entropy, and
the total entropy is the sum  of their contributions.  If, in some
epoch, a single type of entropy makes a dominant contribution to
$S^H$, for example, of the form $S^H= |H|^\alpha$,  $\alpha$ being
a constant characterizing the type of entropy source, and
therefore $S=(a|H|)^3 |H|^\alpha$, eq.(\ref{myeq}) becomes an
explicit inequality,
\begin{equation}
 3H+(3+\alpha)\frac{{\dot H}}{H} \ge 0,
 \label{ec}
\end{equation}
which can be translated into energy conditions constraining the
energy density  $\rho$, and the pressure $p$ of (effective)
sources. Using the Friedman-Robertson-Walker (FRW) equations,
\begin{eqnarray}
 H^2 &=&\ \frac{1}{6}\ \rho \nonumber \\
 {\dot H} &=& -\frac{1}{4} (\rho+p) \label{frw} \\
 {\dot \rho} &+& 3H (\rho+p)=0, \nonumber
\end{eqnarray}
and assuming $\alpha>-3$ (which we will see later is a reasonable
assumption ) and of course $\rho>0$, we obtain
\begin{eqnarray}
 \frac{p}{\rho}&\le&\frac{2}{3+\alpha }-1
 \hspace{.25in} \hbox{for} \hspace{.25in}
 H>0,
 \label{adeqp}\\
 \frac{p}{\rho}&\ge&\frac{2}{3+\alpha }-1
 \hspace{.25in} \hbox{for} \hspace{.25in} H<0.
 \label{adeqn}
\end{eqnarray}
Adiabatic evolution occurs when the inequalities in
eqs.(\ref{adeqp},\ref{adeqn}) are saturated.

A few remarks about the allowed range of values of $\alpha$ are in
order. First, note that the usual adiabatic expansion of a
radiation dominated universe with $p/\rho=1/3$ corresponds to
$\alpha=-3/2$. Adiabatic evolution with $p/\rho<-1$, for which the
null energy condition is violated would require a source for which
$\alpha<-3$. This is problematic since it does not  allow a flat
space limit of vanishing $H$ with finite entropy. The existence of
an entropy source with $\alpha$ in the range $\alpha<-2$ does not
allow a finite $\partial_t S$ in the flat space limit and is
therefore suspected of being unphysical.  Finally, the equation of
state
 $p=-\rho$ (deSitter inflation),  cannot be
described as adiabatic evolution for any finite $\alpha$.

Let us discuss in more detail three specific examples. First, as
already noted, we have verified  that thermal entropy during
radiation dominated (RD) evolution can be described without
difficulties, as expected. In this case, $\alpha=-\frac{3}{2}$,
reproduces the well known adiabatic expansion, but also allows
entropy production. The present era of matter domination requires
a more complicated description since in this case one source
provides the entropy, and another source the energy.

The second case is that of the conjectured
geometric entropy $S_g$, whose source
is the existence of a cosmological horizon
\cite{gibbons,srednicki}. The concept of geometric entropy is
closely related to the holographic principle, and it has appeared
in this connection recently in  discussion of cosmological entropy
bounds. For a system with a cosmological horizon $S^H_g$ is given
by (ignoring numerical factors of order unity)
\begin{equation}
 S_{g}^{H}=|H|^{-2} G_N^{-1}.
 \label{gentropy}
\end{equation}
The equation of state corresponding to adiabatic evolution with
dominant $S_g$, is obtained by substituting $\alpha=-2$ into
eqs.(\ref{adeqp},\ref{adeqn}), leading to $p/\rho=1$ for positive
and negative $H$. This equation of state is simply that of a free
massless scalar field, also recognized as  the two dilaton-driven
inflation (DDI) $(\pm)$ vacuum branches of `pre-big-bang' string
cosmology \cite{pbb} in the Einstein frame.  In \cite{veb} this
was found for the $(+)$ branch in the string frame as an
``empirical" observation. In general, for the case of dominant
geometric entropy, GSL requires, for positive $H$,
$
 p \le \rho,
$
obtained also by \cite{fs} and \cite{EL} using a different
argument. Note that deSitter inflation (DSI) is definitely
allowed. For negative $H$, GSL requires
$
 \rho \le p,
$
and therefore forbids, for example, a time reversed history of our
universe, or a contracting deSitter universe with a negative
constant $H$, unless some additional entropy sources appear.

The third case is that of quantum entropy $S_{q}$, associated with quantum
fluctuations. This form of entropy was discussed in \cite{BMP,gg}. Specific
quantum entropy for a single physical degree of freedom is approximately given by
(again, ignoring numerical factors of order unity)
\begin{equation}
 s_{q}= \int d^3k \ln n_k,
 \label{quantums}
\end{equation}
where $n_k \gg 1$ are occupation numbers of quantum modes \cite{note1}. Note
that quantum entropy  is large for highly excited
quantum states, such as the squeezed states obtained by
amplification of quantum fluctuations during inflation. Quantum
entropy does not seem to be expressible in general as $S^H_{q}=
|H|^{\alpha}$, since occupation numbers depend on the whole
history of the evolution. We will discuss this form of entropy in
more detail later, when the quantum version of GSL is proposed.

We would like to show  that it is possible to formally define a
temperature, and that the definition is compatible with the a
generalized form of the first law of thermodynamics. Recall that
the first law for a closed system states that
$
 T dS = dE+pdV = (\rho+p)dV +V d\rho.
$
Let us now consider the case of single entropy source and formally
define a temperature $T$,
$
 T^{-1}=\left( \frac{\partial S}{\partial E}\right)_V
 =\frac{\partial s}{\partial \rho },
$
since $E=\rho V$ and $S=sV$.  Using eqs.(\ref{frw}), and
$s=|H|^{\alpha+3}$, we obtain
$
 \frac{\partial s}{\partial \rho}=
 \frac{\alpha+3}{12} |H|^{\alpha+1},
$
and therefore
\begin{equation}
 T=\frac{12}{\alpha+3} |H|^{-\alpha-1}.
 \label{gent}
\end{equation}
Note that to ensure  positive temperatures
$
\alpha > -3,
$
a condition which we have already encountered. Note also that
 for $\alpha>-1$,  $T$ diverges in the flat space
 limit, and therefore such a source is
 suspect of being unphysical, leading to the conclusion that the
 physical range of $\alpha$ is $-2\le \alpha\le-1$.
A compatibility check requires
$
 T^{-1}=
 \frac{ \partial s}{\partial t} / \frac{\partial \rho}{\partial t}
$, which indeed yields a result in agreement with (\ref{gent}).
Yet another thermodynamic relation
$
p/T=\left( \frac{\partial S}{\partial V}\right)_E$, leads to
$p=sT-\rho$ and therefore to
$
 p/\rho =\frac{2}{\alpha+3}-1
$
for adiabatic evolution, in complete agreement with
eqs.(\ref{adeqp},\ref{adeqn}). For $\alpha=-2$, eq.(\ref{gent})
implies $T_g=|H|$, in agreement with \cite{gibbons},  and  for
ordinary thermal entropy $\alpha=-3/2$ reproduces the known
result, $T=|H|^{1/2}$.

We turn now to discuss entropy bounds, GSL and cosmological singularities.
First, we discuss compatibility of entropy bounds and GSL, and then use GSL to
derive a new bound relevant to cosmological singularities. Bekenstein
\cite{beb2} suggested that in flat space there is a universal entropy bound on
the maximal entropy content in a region containing energy $E$ and of size $L$,
$S<E L$, and then applied this idea to cosmology \cite{beb1}, by choosing the
particle horizon $ d_p= a(t)\int \frac{dt'}{a(t')}$ as $L$. Recently Veneziano
\cite{veb} argued that since a black hole larger than the horizon cannot form,
the largest entropy in a region corresponds to having just one black hole per
Hubble volume $H^{-3}$, namely (introducing the Planck mass $M_p=G_N^{-1/2}$)
that $s  \le M_p^2 |H|$ and
\begin{equation}
S^H \le M_p^2 |H|^{-2}.
 \label{VEB}
\end{equation}
This conjecture was further supported in \cite{KL}.
Perhaps a link between the two distinct entropy bounds can be
established by choosing instead of $d_p$, the Hubble radius
$H^{-1}$ \cite{veb,EL,KL} and since $E_H=M_p^2 |H|^{-1}$, the
condition $S< E L$ is translated into eq.(\ref{VEB}). Note that
when applied to non-inflationary cosmology, as done in
\cite{beb1}, particle horizon and Hubble radius are about the same
and therefore both bounds give similar constraints on $S^H$.
A consequence of bound (\ref{VEB}) is therefore that geometric entropy should
always be the dominant source of entropy,
\begin{equation}
 S^H \le  S_g^H.
 \label{genbeb}
\end{equation}
In \cite{KL} an example of an expanding and recontracting universe with some
matter and a small negative cosmological constant was presented, for which
bound (\ref{VEB}) seems to be violated. This example involves an epoch in which
the causal range is very different from $|H|^{-1}$, and is quite interesting,
but its resolution will not affect our conclusions for the cases we are
interested in, in which $H^2$ is at least as large as $|\dot H|$.

Is GSL compatible with entropy bounds? Let us start answering this question by
considering a universe undergoing decelerated expansion, that is  $H>0$, $\dot
H<0$. For entropy sources with $\alpha>-2$, going backwards in time, $H$ is
prevented by the entropy bound (\ref{genbeb}) from becoming too large. This
requires that at a certain moment in time $\dot H$ has reversed sign, or at
least vanished. GSL allows such a transition. Evolving from the past towards
the future, and looking at eq.(\ref{ec}) we see that  a transition from an
epoch of accelerated expansion $H>0$, $\dot H >0$, to an epoch of decelerated
expansion $H>0$, $\dot H <0$, can occur without violation of GSL. But later we
discuss a new bound appearing in this situation when quantum effects are
included.

For a contracting universe with $H<0$, and if sources with $\alpha>-2$ exist,
the situation is more interesting. Let us check whether in an epoch of
accelerated contraction   $H<0$, $\dot H <0$, GSL is compatible with entropy
bounds. If an epoch of accelerated contraction lasts, it will inevitably run
into a future singularity, in conflict with  bound (\ref{genbeb}). This
conflict could perhaps have been prevented if at some moment in time the
evolution had turned into decelerated contraction with $H<0$, $\dot H > 0$. But
a brief look at eq.(\ref{ec}), $\dot H \le -\frac{3}{3+\alpha} H^2$, shows that
decelerated contraction is not allowed by GSL.  The conclusion is that for the
case of accelerated contraction GSL and the entropy bound are not compatible.

To resolve the conflict between GSL and the entropy bound, we
propose adding a missing quantum entropy term
$
 dS_{Quantum}=  - \mu dn_H,
$
where $\mu(a,H,\dot H,...)$ is a ``chemical potential" motivated by the
following heuristic argument. Specific quantum entropy is given by
(\ref{quantums}), and we consider for the moment one type of quantum
fluctuations that preserves its identity throughout the evolution. Changes in
$S_q$ result from the well known phenomenon of freezing and defreezing of
quantum fluctuations. For example, quantum modes whose wavelength is stretched
by an accelerated cosmic expansion to the point that it is larger than the
horizon, become frozen (``exit the horizon"), and are lost as dynamical modes,
and conversely quantum modes whose wavelength shrinks during a period of
decelerated expansion (``reenter the horizon"), thaw and become dynamical
again. Taking into account this ``quantum leakage" of entropy, requires that
the first law should be modified as in open systems $TdS=dE+PdV-\mu dN$, as
first suggested in \cite{prigogine}.

In a universe going through a period of decelerated
expansion, containing some quantum fluctuations which have
reentered the horizon (e.g., a homogeneous and isotropic 
background of gravitational waves), physical 
momenta simply redshift, but since
no new modes have reentered, and since occupation numbers do not
change by simple redshift, then within a fixed comoving volume,
entropy does not change. However, if there are some frozen
fluctuations outside the horizon ``waiting to reenter" then there
will be a change in quantum entropy, because the minimal comoving
wave number of  dynamical modes $k_{min}$, will decrease due to
the expansion, $k_{min}(t+\delta t)<k_{min}(t)$. The resulting
change in quantum entropy, for a single physical degree of freedom, is
$
 \Delta s_{q}=\!\!\!
  \int\limits_{k_{min}(t+\delta t)}^{k_{min}(t)}\!\!\!
  k^2 dk \ln n_k,
  $
 and since $k_{min}(t)=a(t) H(t)$,
$
 \Delta S_{q}= \!\!\! \int\limits_{a(t+\delta t)H(t+\delta t)}^{a(t) H(t)}
  k^2 dk \ln n_k =
 -  \Delta (aH)^3 \ln n_{k=aH},
$
provided $\ln n_k$ is a smooth enough function. Therefore, for $N$ physical
degrees of freedom, and since
$n_H=(aH)^3$,
\begin{equation}
 d S_{q} = - \mu N d n_H,
 \label{conjecture}
\end{equation}
where parameter $\mu$ is taken to be positive. Obviously, the
result depends on the spectrum $n_k$, but typical spectra are of
the form $n_k\sim k^\beta$, and therefore we may take as a
reasonable approximation  $\ln n_k\sim constant$ for all $N$ physical degrees of
freedom.

We adopt proposal (\ref{conjecture}) in  general,
\begin{eqnarray}
 dS &=& dS_{Classical}+dS_{Quantum} \nonumber \\
  &=&dn_H S^H+ n_H dS^H- \mu N dn_H,
 \label{q2ndlaw}
\end{eqnarray}
where $S^H$ is the classical entropy within a cosmological
horizon.
 In particular, for the case that $S^H$ is dominated by a single
 source $S^H=|H|^\alpha$,
\begin{equation}
 \left(3H+  3  \frac{{\dot H}}{H}\right) n_H (S^H-  \mu N) +
 \alpha \frac{{\dot H}}{H} n_H S^H  \ge 0.
\label{qec}
\end{equation}

Quantum modified GSL (\ref{qec}) allows a transition from accelerated to
decelerated contraction. As a check, look at $H<0$, $\dot H=0$, in this case
modified GSL requires $ 3H (S^H-\mu N) \ge 0, $ which, if $\mu N \ge S^H $, is
allowed.  If the dominant form of entropy is indeed geometric entropy, the
transition from accelerated to decelerated contraction is allowed already at
$|H|\sim M_p/ \sqrt{N}$. In models where $N$ is a large number, such as grand
unified theories and string theory where it is expected to be  of the order of
1000, the transition  can occur  at
a scale much below the Planck scale, at which classical general relativity is
conventionally expected to adequately describe background evolution.

If we reconsider the transition from accelerated  to decelerated expansion
and require that (\ref{qec}) holds, we discover a
new bound derived directly from GSL, compatible with,  but not  relying on, 
bound (\ref{genbeb}). Consider the case in which $\dot H$ and $H$ are
positive, or $H$ positive and $\dot H$ negative but $|\dot H|\ll H^2$, relevant
to whether the transition is allowed by GSL. In this case, (\ref{qec}) reduces
to $S^H-\mu N\ge 0$, that is, GSL puts a lower bound on the classical entropy
within the horizon. If geometric entropy is the dominant source of entropy as
expected,  GSL puts a lower bound on geometric
entropy $S_g^H\ge \mu N$, which yields an upper bound on $H$,
\begin{equation}
 H\le \frac{M_p}{\sqrt{N}}.
\label{upperH}
\end{equation}
The scale that appeared previously in the resolution of the
conflict between entropy bounds and GSL for a contracting universe has
reappeared in (\ref{upperH}), and remarkably, (\ref{upperH}) is the same bound
obtained in \cite{beb1} using different arguments. Bound (\ref{upperH}) forbids
a large class of singular homogeneous, isotropic, spatially flat cosmologies by
bounding their curvature.

 An interesting study case  is `pre-big-bang' string cosmology \cite{pbb}.
In this scenario the evolution of the universe starts
from a state of very small curvature and string coupling,
undergoes a phase of dilaton-driven inflation (DDI) which
 joins smoothly standard radiation dominated (RD) cosmology, 
thus giving rise
to a singularity free inflationary cosmology. The graceful exit transition
from DDI to  RD, has been studied intensely \cite{ge}, with the
following scenario emerging, first classical corrections limit the
curvature by trapping the universe in an  algebraic fixed point
\cite{gmv} a linear dilaton deSitter solution, and then quantum
corrections limit the string coupling, and end the transition
\cite{bm,bmuv}. Modified GSL supports this exit scenario,
 clarifies the conditions for
the existence of the algebraic fixed point, determines new energy
conditions, and constrains sources required to complete a graceful
exit transition.

We present here the case of dominant $S_g$.
A candidate geometric entropy is  given by the analog of
 eq.(\ref{gentropy}) \cite{veb} by substituting $M_p^2=e^{-\phi}M_S$,
 $M_S$ being the (constant) string mass and $\phi$ is the dilaton,
 $S^H=e^{-\phi} H^{-2}$. The expression for $n_H$
is unchanged. Condition (\ref{ec})
now reads
$
 3H+ \frac{{\dot H}}{H}-\dot\phi \ge 0.
$
Using $\pbd=\dot\phi-3H$, we obtain
$
 \frac{{\dot H}}{H}-\pbd \ge 0,
$
leading, by using one of string cosmology equations of motion
$\bar \sigma-2 \dot H+2 H \pbd = 0$ \cite{bm3}, for $H>0$, which
is the natural choice for pre-big-bang phase, to the energy condition
$
 \bar\sigma \ge 0
$ \cite{dick}. An immediate consequence  is that if $\dot H$
vanishes, then $\pbd<0$, so an algebraic fixed point \cite{gmv}
necessarily has to occur for $\pbd<0$. The same conclusion was
reached in \cite{veb}, and previously in \cite{gmv,bm}. Further
investigation is required  clarify
the correct comparison to the analysis in ordinary FRW cosmology.

\acknowledgements
Work  supported in part by the  Israel Science Foundation. It is a pleasure to
thank R. Madden and G. Veneziano for useful discussions and helpful suggestions
and comments,   S. Foffa and R. Sturani for discussions about string cosmology,
D. Eichler and J. Donoghue for discussions, and J. Bekenstein, R. Easther, N.
Kaloper and A. Linde for comments on the manuscript.

\end{document}